# Evaluation of Cyber Attacks Targeting Internet Facing IoT: An Experimental Evaluation

Navod Neranjan Thilakrathne [1,*] , Dr. Rohan Samarasinghe [2] , Madhuka Priyashan[3]

[1,2]Department of ICT, Faculty of Technology, University of Colombo

[3] Department of Mechanical and Manufacturing Engineering, Faculty of Engineering, University of Ruhuna, Galle, Sri Lanka

navod.neranjan@ict.cmb.ac.lk*

**Abstract:** The rapid growth of Information and Communication Technology (ICT) in the 21st century has resulted in the emergence of a novel technological paradigm; known as the Internet of Things, or IoT. The IoT, which is at the heart of today's smart infrastructure, aids in the creation of a ubiquitous network of things by simplifying interconnection between smart digital devices and enabling Machine to Machine (M2M) communication. As of now, there are numerous examples of IoT use cases available, assisting every person in this world towards making their lives easier and more convenient. With the latest advancement of IoT in variety of domains such as healthcare, smart city, smart agriculture it has led to an exponential growth of cyber-attacks that targets these pervasive IoT environments, which can even lead to jeopardizing the lives of peoples; that are involving with it. In general, this IoT can be considered as every digital object that is connected to the Internet for intercommunication. Hence in this regard in order to analyse cyber threats that come through the Internet, here we are doing an experimental evaluation to analyse the requests, received to exploit the opened Secure Shell (SSH) connection service of an IoT device,  which in our case a Raspberry Pi devices, which connected to the Internet for more than six consecutive days. By opening the SSH service on Raspberry Pi, it acts as a Honeypot device where we can log and retrieve all login attempt requests received to the SSH service opened. Inspired by evaluating the IoT security attacks that target objects in the pervasive IoT environment, after retrieving all the login requests that made through the open SSH connection we then provide a comprehensive analysis along with our observations about the origin of the requests  and the focus areas of intruders ; in this study.

**Keywords:** IoT, Cyber-attack, Honeypot, Cyber security, Internet security

## 1. Background

With the rapid advancement of IoT, it is being widely integrated and applied in with, many application domains for instance military, agriculture, smarty city, surveillance, healthcare, and so on, offering a wide variety of benefits for the goodness of mankind. This recent advancement of IoT, made it a fast-growing technology conquering all other technologies present, creating a huge impact on the society as well as many of the business operations [1]-[5]. The IoT has gradually permeated most of the aspects of human life such as medical care, agriculture, military, and so on, making the technology affordable for everyone [6]-[10]. In recent years, owing to the rapid demand and various socio-economic factors the number of IoT devices, that are commissioned has shown an exponential growth making this IoT is a part of our daily life. On the other hand, this also poses another doubt about robust security approaches in response to this rapid demand over the years, as the number of cyber-attacks that target digital infrastructure becomes more robust, intuitive, and complex owing to the vast threat landscape. Not only that but it is also evident that the number of potential attackers is also increasing with novel sophisticated tools and techniques, which clearly necessitates the implementation of appropriate security measures [1],[10]-[15]. Inspired by analyzing these IoT security attacks that target digital objects in the pervasive IoT environment, in this study we do an experimental evaluation to set up a Honeypot IoT devices using Raspberry Pi devices along with open source technologies and we will keep open the Secure Shell (SSH) connection service of Raspberry Pi for more than six consecutive days to log all the requests, that is received via the open SSH service, that are coming to the internal network, from the outside Internet.

When it comes to a Honeypot it looks and functions exactly like a genuine computer system, including all programs and data, deceiving hackers into believing it as a legitimate target. A honeypot, for example, may imitate a company's customer billing system, which is a common target for intruders looking for credit card data [1]-[3]. Once the hackers have gained access, their activities may be recorded and analysed for clues on how to make the real network more secure and resilient which is highly beneficial for further research and security product development. On the other hand, honeypots are made appealing to attackers by including intentional security flaws, which is the exact theory that our research is built upon. For example, it may include ports that respond to a port scan or weak passwords. Then to attract attackers into the honeypot environment, vulnerable ports may be left open so attackers will lure into the honeypot instead of the real system [18]-[20]. By monitoring traffic coming into the honeypot system, we can assess; where the cybercriminals are coming from (01), the level of threat (02), what data or applications they are interested in (03), and how well the security measures are working to against towards stopping cybercriminals (04).

Based on the literature it is evident that; there are already studies available about this honeypot where several reviews /surveys have also been done [3]-[10]. On the other hand variety of research has also been carried out about implementing honeypot systems and then using artificial intelligence towards identifying threats. But in our study, we intend to use a cost-effective devices and open source freely available technologies towards making an IoT honeypot device to assess the threats and derive conclusions for further research [15]-[20]. Nevertheless, in the literature; there was less research has done about IoT honeypot architecture and assessments hence we believe our research experiment would be useful for carrying out further research in this area.

In order to represent our research work the rest of the paper is organized as follows. Section two (02) describes the objectives of our study and section (03) depict our methodology. After that in section four (04), we highlight our results and finally, we conclude our paper with the conclusion section.

## 2. Objectives

The main objective of our research is to analyse the cyber-attacks that target typical IoT objects in the pervasive IoT environment using a really simple cost-effective setup, and have an exact idea about what these external intruders are trying to do along with their origin source, when it comes to exploiting a remote IoT device, in a typical IoT network; and then suggest recommendations for stopping these threats and future research.

## 3. Methodology

For setting up the research environment; configuration and the environmental setup were done and it comprised of four (04) key steps as follows.

1. *First, for setting up the IoT honeypot device, in an internet-facing network; a Raspberry Pi derive is used which has the pre-configured Raspberry Pi OS.*
2. *Secondly in order to make our Raspberry Pi module a honeypot; an open-source SSH Pluggable Authentication Module (PAM) is set up on the device (For setting the authentication of OS), which is an open source software package freely available on GitHub.*
3. *Next after implementing the PAM module an open-source portable OpenSSH is configured on the Raspberry Pi device, which is also an open source software package freely available on GitHub. (This OpenSSH is a complete implementation of the SSH protocol (version 2) for secure remote login, command execution, and file transfer)*
4. *Then upon the completion of all these configurations, all the login requests that may receive from external parties are configured to log in to the PAM database by default which is in the Raspberry device, where we can use them for our analysis task using a Microsoft SQL server database collating them into a single CSV file, afterward.*
5. *For better understanding the Fig. 1 depicts the high-level architecture of our honeypot setup.*

After setting all these configurations, the SSH service which is set up on the Raspberry Pi kept open all the time. Then we connected the Raspberry Pi device with the internet via a broadband router and it was the sole device that is connected to the Internet. The Internet connection itself was a standard residential broadband ADSL service. Finally, the configured IoT SSH honeypot was kept online for over six (06) days of time and afterward logs were taken out and imported to the local computers Microsoft SQL server database for further analysis.

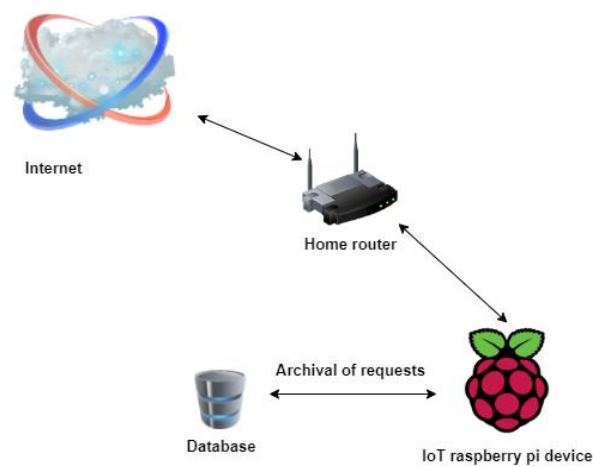

**Fig. 1** The IoT Honeypot architecture

## 4. Results

The SSH service which is set up on the Raspberry Pi was running all the time and was the sole device on the Internet connection. The Internet connection itself was a standard residential broadband ADSL service. Finally, the configured IoT SSH honeypot was kept online for over six (06) days of time. During the time we witnessed a total of **105764** logging attempts. While further breaking down all these requests; the interesting thing what we noted was the vast majority of requests came from **four (04)** major hosts (sources), which accounted for **90175** requests on the whole. (around 85% from all the requests). The Fig. 2 depicts the analysis based on the dispersion of the requests based on geographic area.

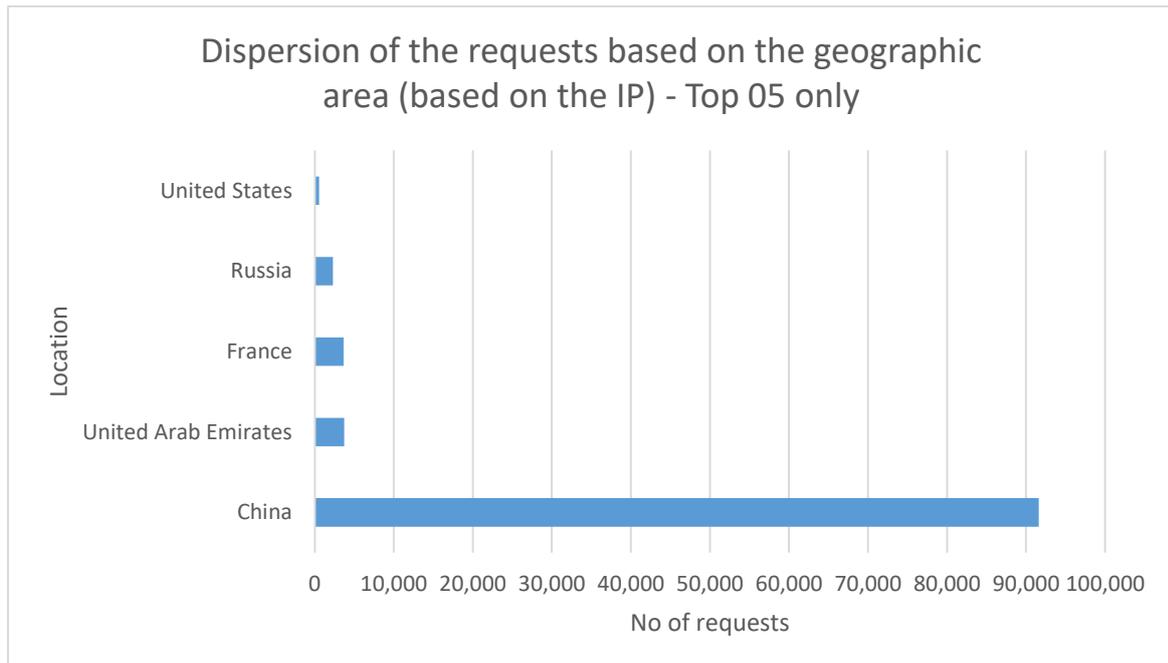

**Fig. 2** Dispersion of the requests based on the geographic area (based on the IP) - Top 05 only

Most of these hosts were attempting to exploit the credentials and they were carrying out a dictionary attack on the root user. The following is a list of the top 20 users that were attempted in order to exploit the login via the open SSH service.

- root -  101634
- admin - 967
- 0 - 98
- pi - 87
- test - 76
- user - 74
- support - 33
- odroid - 28
- ftpuser - 26
- postgres - 25
- ubnt - 24
- guest - 23
- ubuntu - 18

- Administrator - 16
- ftp - 14
- oracle - 14
- vagrant - 14
- webmaster - 14
- administrator - 13
- dietpi - 13

Then in the following, we depict the passwords that were attempted. The following is the list of the top 20 passwords we have noted.

- admin
- 0
- 123456
- password
- 1234
- raspberry
- root
- ubnt
- administrator
- welc0me
- guest
- openelec
- test
- pi
- alpine
- default
- 12345
- postgres
- alex
- ftpuser

Based on the results of this simple experiment, following are the recommendations and conclusions that we have derived from our research.

- It is evident that most of the intruder sources are dispersed around China which accounts for more than 85% of the exploitation attacks, which we can clearly say most of the attackers are located in China which we needed to be aware of.
- Followed by China we have the United Arab Emirates, France, Russia and Unites states in the top list, which these malicious hosts are based on; whereas the number of requests received to our honeypot were negligible from these sources as opposed to the amount of requests that were received from China.
- Based on most of the dictionary attacks performed it is evident that most of the attackers are trying to exploit the default user names or passwords (factory fitted username and password). For instance, the default username and password for Kali Linux distribution are root and toor. For some other distributions, it is admin and admin for both username and password.

More examples (See the Table 1 below):

Table 1. More examples for default credentials of IoT devices

| Username | Password |
|----------|----------|
| admin | password |
| admin | 12345 |
| guest | password |
| guest | 12345 |

So we can clearly assume all these external attacks that target these IoT devices are trying to find a simple vulnerability from the end-users end. (e.g.: not changing the default user name and password and keep open the device connection for the internet), where it would be the easiest way for attackers to exploit your IoT devices for planting malware, used in a botnet for performing Distributed Denial of Service (DDOS) attacks, or performing other types of malicious attacks. Hence based on these conclusions what we can recommend is, it is always advisable to use a reliable antivirus in your personal commuter or smart mobile device .where it would hinder the path of attackers towards infecting your device. On the other hand end users needed to be vigilant for their operating system, inbuilt firewall activities as some software's may change the configurations on these firewall settings especially with Microsoft window 8, 8.1, and 10 operating systems, as even your smart mobile device or personal computer can act as an IoT device. On the other hand, when it comes to other types of IoT devices such as smart TV, webcams, surveillance cameras, and routers it is more advisable to change the factory fitted default user names and passwords as it would make it super easy for attackers to intrude into the system. At last but not least whenever you are connecting with the internet it is advisable to think that you are one step close towards getting hacked.

**Recommendations**

a) For policy makers and planners

Protecting user privacy is a vital element nowadays owing to the increasing cyber threats that come through the Internet. Whenever a user connects an IoT device to the Internet if the device is not safe or if the user is unaware of the security and privacy of themselves it would make a perfect gap for cybercriminals to access. Hence the security software developers and relevant stakeholders can take appropriate actions towards protecting their user's privacy even the user is not aware about any of the things.

b) For researchers

In this study we have made a simple and a cost effective IoT honeypot architecture for analysing threats that target IoT devices in the pervasive IoT devices and implemented it by connecting to the Internet. Hence this would make a good avenue for future researchers to carry out further research making our study as a foundation as there are less research has been carried out targeting IoT ecosystem.

## 5. Conclusion

Inspired by evaluating the IoT security attacks that target objects in the pervasive IoT environment, we have configured a simple cost effective IoT honeypot architecture and we then provide a comprehensive analysis of requests received along with our observations and recommendations; in this study. Based on our analysis it is evident that leaving services such as SSH open can result in a large number of malicious requests. Hence the users need to be sure to lock down any ports you do not require; from the firewall regardless of their operating system. All these external attacks that target these IoT devices are trying to finding a simple vulnerability and if the end-users are not aware of any of the things (not changing the default user name and password and keep open the connection for the internet), it would be the easiest way to attackers to exploit your IoT devices such as your smart watch, desktop computer, and laptop device and etc. Thus we believed this study would be useful for anyone, who seeks to protect their digital privacy. On the other hand in this study, we have showcased a simple IoT honeypot architecture and demonstrated its ability by connecting with the real Internet, where this would be highly useful for carrying out further research.